\documentclass[aps,prl]{revtex4}
\def\ee{\end{equation}}
\def\bea{\begin{eqnarray}}

\def\bra#1{\langle #1 |}
\def\ket#1{| #1\rangle}

\def\Tr{{\rm Tr}}

\usepackage{graphicx}
\DeclareGraphicsExtensions{.png}
\begin{document}

\title{Simple Refutation of the Eppley-Hannah argument} 
\author{ Adrian Kent}
\email{A.P.A.Kent@damtp.cam.ac.uk} 
\affiliation{Centre for Quantum Information and Foundations, DAMTP, Centre for
Mathematical Sciences,
University of Cambridge, Wilberforce Road, Cambridge, CB3 0WA, United Kingdom}
\affiliation{Perimeter Institute for Theoretical Physics, 31 Caroline Street
North, Waterloo, ON N2L 2Y5, Canada.}

\begin{abstract}
In an influential paper, Eppley and Hannah argued that gravity must
necessarily be quantized, by proposing a thought experiment 
involving classical gravitational waves 
interacting with quantum matter.
They argue the interaction must either violate
the uncertainty principle or allow superluminal signalling.  
The feasibility of implementing their experiment in
our universe has been challenged by Mattingly, 
and other limitations of the argument have been noted 
by Huggett and Callender and by Albers et al.. 
However, these critiques do not directly refute the
claim that coupling quantum theories with a Copenhagen collapse
postulate to unentanglable classical gravitational degrees of freedom
leads to contradiction. 
I note here that if the gravitational
field interacts with matter via the local quantum state, the Eppley-Hannah 
argument evidently fails.   This seems a plausibly natural feature 
of a hybrid theory, whereas the alternative considered by
Eppley-Hannah is evidently inconsistent with relativity.
\end{abstract}
\maketitle
\section{Introduction}

In a paper \cite{eppley1977necessity} that has had significant
influence
in the quantum gravity community, Eppley and Hannah argued for the necessity of 
quantizing gravity, by claiming to show that a contradiction would
arise in any dynamical theory in which a classical gravitational field
interacts with quantum matter.   Their argument invokes a 
thought experiment, in which the experimenter prepares gravitational
waves of very short wavelength and low momentum (a consistent
combination in classical theories) and uses these to probe
the state of a localized quantum particle.  
They summarize their argument thus: 
\begin{quote}
Briefly, we show that if a gravitational wave of arbitrarily small 
momentum can be used to make a position measurement on a quantum particle,
i.e., to ``collapse the wave function into an eigenstate of
position,'' then the
uncertainty principle is violated. If the interaction does not result in collapse
of the wave function, it is then possible to distinguish experimentally between
superposition states and eigenstates. We show that this ability allows one to
send observable signals faster than $c$ when applied to a state consisting of two
spatially separated particles with correlated spins.
\end{quote}

Huggett and Callender \cite{huggett2001quantize} noted that Eppley and Hannah's arguments
implicitly assume some version of quantum theory in which a collapse
postulate applies, at least to standard 
measurements by standard matter-based devices.\footnote{As we discuss
below, Eppley and Hannah suppose that the gravitational field can also
be used to carry out measurements on quantum systems, and that 
these measurements need not necessarily involve collapse.}
It is not clear, for example, how to adapt
their arguments to a theory in which matter is described
by Everettian unitary quantum theory while the gravitational 
field is described classically.   Similar comments apply
to a de Broglie-Bohm quantum theory of matter coupled to
classical gravity.      

That said, versions of Copenhagen 
quantum theory with collapse postulates remain 
popular and may possibly be at least correct enough
to justify Eppley-Hannah's use of a projection postulate
(though not, as we explain below, their inconsistent
application of the projection postulate in relativistic
quantum theory).   The many versions  \cite{saunders2010many} of
Everettian quantum theory all have their own
difficulties \cite{saunders2010many}.
The most straightforward way of trying to couple Everettian
quantum theory to classical gravity, via semi-classical gravity
theories, is inconsistent with observation \cite{page1981indirect}.
The same difficulty holds if a classical gravitational field is coupled
to the unitarily evolving wave function in a de Broglie-Bohm
theory of matter, while coupling a classical gravitational field to 
the de  Broglie-Bohm particles seems likely to allow superluminal
signalling.    
Huggett and Callender's criticisms thus qualify the scope of the
Eppley-Hannah argument, but (depending on one's perspective on
quantum foundations) may not necessarily remove much of its force.

Mattingly \cite{mattingly2006eppley} challenged the argument on different grounds, 
arguing in impressive detail that the thought experiment is not merely
impossible with any foreseeable technology but likely impossible
in principle, given the known values of fundamental constants, 
the size of the observable universe and the amount of matter
it contains.   This makes the thought experiment's status 
questionable: what logical force does it have if physics 
makes it impossible?    

Here again, there is room for debate.   Mattingly did not claim
to show that no experiment along the lines of Eppley and Hannah's
proposal could ever be carried out, only that their specific
proposal could not.   The burden of proof may indeed be, as
he noted, with Eppley and Hannah, but some may 
nonetheless have felt encouraged to think that some feasible
experiment could be devised with more ingenuity, even if 
a concrete proposal presently eludes them. 
A perhaps stronger counter-argument is that Eppley and Hannah's 
thought experiment seems to be pointing to a conceptual problem preventing a 
unified theory of classical gravity and quantum matter.
On one view of physics, if this problem is solvable, it ought to 
be solvable in a large class of logically possible universes, 
including universes that do not have the apparently contingent features
that preclude carrying out the thought experiment in ours. 
On this view, if the thought experiment leads to contradiction in such
universes, the problem is (most likely) unsolvable.   

Albers, Kiefer and Reginatto (AKR) \cite{albers2008measurement} noted a key error in Eppley-Hannah's
discussion of entangled systems and presented a model of a 
gravitational wave interacting with a quantized scalar field.
AKR's model is based on N\"ordstrom's scalar theory of gravity 
in $1+1$ dimensional Minkowski space, rather than a standard theory
in $3+1$ dimensions. This model was chosen to simplify
the calculations.  Their discussion is nonetheless still quite complex, and
solutions for the relevant interacting case are presented only
to first order in $g$, and are argued to be valid to this order
only for specific quantum field states and gravitational waves
satisfying particular conditions.   Moreover, the formalism
of interacting classical-quantum systems they consider
leads them to conclude that ``a consistent theory of 
interacting classical-quantum systems leads to final
states that are entangled''.   To be clear, this entanglement
is meant to be {\it between} the classical and quantum sectors. 

Clearly, there is no universal consensus on precisely
what it means to say a theory is classical, or on
how to delineate the class of classical-quantum hybrid theories.
Nonetheless, from a modern information-theoretic perspective on physics,
this conclusion seems misstated: a key difference between
classical and quantum degrees of freedom is that classical 
degrees of freedom cannot be entangled with anything.   
The discussion of Ref. \cite{albers2008measurement} thus does not seem to apply
to classical-quantum hybrid theories as most theorists
would currently understand the term.  

No doubt, these counter-arguments also have counters. 
The net result, though, is that the Eppley-Hannah argument
continues to be influential and is seen as a relevant, albeit 
perhaps not decisive, reason for favouring quantum gravity over hybrid
theories.  (See, for example, Refs. \cite{kiefer2012quantum,hossenfelder2012eppley,sep-quantum-gravity} for 
recent discussions.)  

\section{Eppley-Hannah and collapse} 

In fact, Eppley-Hannah's argument is flawed, even granting the
assumption of a version of quantum theory with collapses, and
regardless of the feasbility of their thought experiment.  

To eliminate their contradiction, it is enough to show that 
one of the alternatives they consider leads to no contradiction.
I will focus on their case $3$, in which they consider ``scattering of the
gravitational wave from the wave function of the quantum particle with no
collapse''. 
As they say, this could in principle be ``an experimental method for observing the wave
function without collapsing it''.  
It could also imply that ``there exists a direct way of viewing [a]
collapse event when it is produced by an ordinary measurement.''
But it is not necessarily the case that this ``would
lead to the possibility of sending signals faster than $c$''
in a hybrid theory, as they argue. 

Eppley-Hannah present an experiment involving entangled photons
split at each end by calcite crystals into polarization-dependent
beams.   They represent the individual photon states by
position space wave functions.   This may concern some readers,
since the status of photon wave functions is controversial.
However, it is not essential to their argument, which could
equally well be carried out by considering entangled states of 
massive particles. 

Eppley-Hannah proceed to argue as follows. 
Suppose we prepare an EPR-type state 
of the form 
\begin{equation} \label{ehepr}
 \psi^L (x-x_0^L ) \psi^R ( x - x_0^R ) + 
 \psi^L (x-x_1^L ) \psi^R ( x - x_1^R ) \, , 
\end{equation} 
where the state $\psi^L ( x - x_0^L )$ is localized
around $x_0^L$ and so on.   
Suppose these factors of the wave function components
are all localized within a scale $a$, so that
\begin{equation}
\int_{ |x| \leq a}  | \psi^L (x) |^2 dx  = 1 \, ,
\end{equation}
and similarly for $\psi^R$.  
Suppose also that $ | x_0^L - x_1^L | = | x_0^R - x_1^R | = b$
where $b>2a$, and that $ | x_0^L - x_0^R | \gg b$. 

Now Eppley-Hannah claim, oddly, that if no measurement has been carried out
on L then the R system is described by the wave function
$\psi^R ( x - x_0^R) + \psi^R ( x - x_1^R )$.  
There are two things wrong with this.  First, one needs to
fix a reference frame in order to relate two time-dependent
statements.   A correct statement would be that, if we calculate and apply
the collapse postulate in a given frame, and if system L has not been measured at time
$t$ in that frame, then in
the standard relativistic quantum formalism we may say the 
wave function at time $t$ in that frame has not collapsed. 
Second, as Albers et al. note \cite{albers2008measurement},
even given such a frame choice, this does not give us a pure state wave function 
for system R.    The correct statement is that system R 
is in an improper mixed state, described by the density matrix
giving a uniform mixture of $\psi^R ( x - x_0^R)$ and  $\psi^R ( x -
x_1^R )$.   However, correcting this statement alone does 
not {\it per se} refute the next part of Eppley-Hannah's argument, since 
this mixture is distinguishable from the individual states
{\it if} one is given a system described by one of the three
possible states and is able (by some hypothetical
classical-quantum probe that does not follow standard
quantum measurement rules) to identify the system's 
quantum state.  

Eppley-Hannah also claim that if a measurement has been
carried out on L, effectively localizing the position
around either $x_0^L$ or $x_1^L$, then the R system is
described either by $\psi^R ( x - x_0^R)$ or by $\psi^R ( x - x_1^R
)$.  
Again, one needs to fix a reference frame to make such a 
statement, although it is true if a suitable frame is
chosen.   In standard relativistic quantum theory, 
of course, while the statement is mathematically true in a suitable frame, it 
is knowable only by an observer who knows that a localised position
measurement has been carried out on $L$.   In order to know which
description of the $R$ system applies, the observer also needs to 
know the outcome of the measurement on the $L$ system.
Both of these are possible for an observer at $L$ but not
(until a time in the future light cone of the measurement at $L$)
for an observer at $R$. 
This frame-dependent
description of the wave function does not lead to superluminal
signalling or other observable consequences in standard relativistic
quantum theory, because the 
wave function is not directly observable and the measurement
postulates ensure that the observable information is 
frame-independent.   

Finally, Eppley-Hannah claim that in a hypothetical hybrid theory
in which interactions between gravitational waves and quantum
matter do not collapse the wave function of the latter,
an interaction between a
classical gravitational wave and system R allows the
first case (no measurement on L) to be distinguished 
from the second case (measurement on L).   

The problem here is that, even if we restrict to Minkowski
space rather than general space-times, it makes no sense to postulate a 
relativistic theory of quantum states, together with a
collapse postulate, in which (i) collapse is a real physical
phenomenon (since position space wave functions have objective and
measurable meaning in the type of hybrid theory that
Eppley-Hannah postulate), and (ii) collapse propagates
instantaneously.   In a relativistic theory instantaneity is
frame-dependent, whereas a real physical effect cannot be.    
The fundamental problem with
Eppley-Hannah's proposal is thus not that it gives rise
to superluminal signalling (which, although problematic 
and counter-intuitive in a relativistic theory, need not necessarily lead to
contradiction (see e.g. Ref. \cite{kent1998causality})), but that it is incoherent. 

There is, however, a consistent way of defining a 
relativistic theory of quantum states, together with
a collapse postulate, in which collapse is a real
physical phenomenon, local quantities 
defined by quantum states have
objective and measurable meaning, and no superluminal
signalling is possible \cite{kent2005nonlinearity,kent2005causal}.
This is to suppose that the effects of collapse propagate
causally, and that the objective and measurable quantity
at a point $x$ is the local quantum state $\rho^{\rm loc}(x)$.

We define $\rho^{\rm loc} (x)$ to be local density matrix of $\psi_{\Lambda(x)}$ at $x$,
taken as the limit of the local
density matrices for the wave functions of spacelike
hypersurfaces $S$ tending to the past light cone $C=\Lambda (x)$
of $x$. 
To understand the definition of $\rho^{\rm loc}(x)$, consider Figure
1, adapted from Ref. \cite{kent2005nonlinearity}. 
\begin{figure*}
\label{one}
\includegraphics[width=4in,keepaspectratio=true]{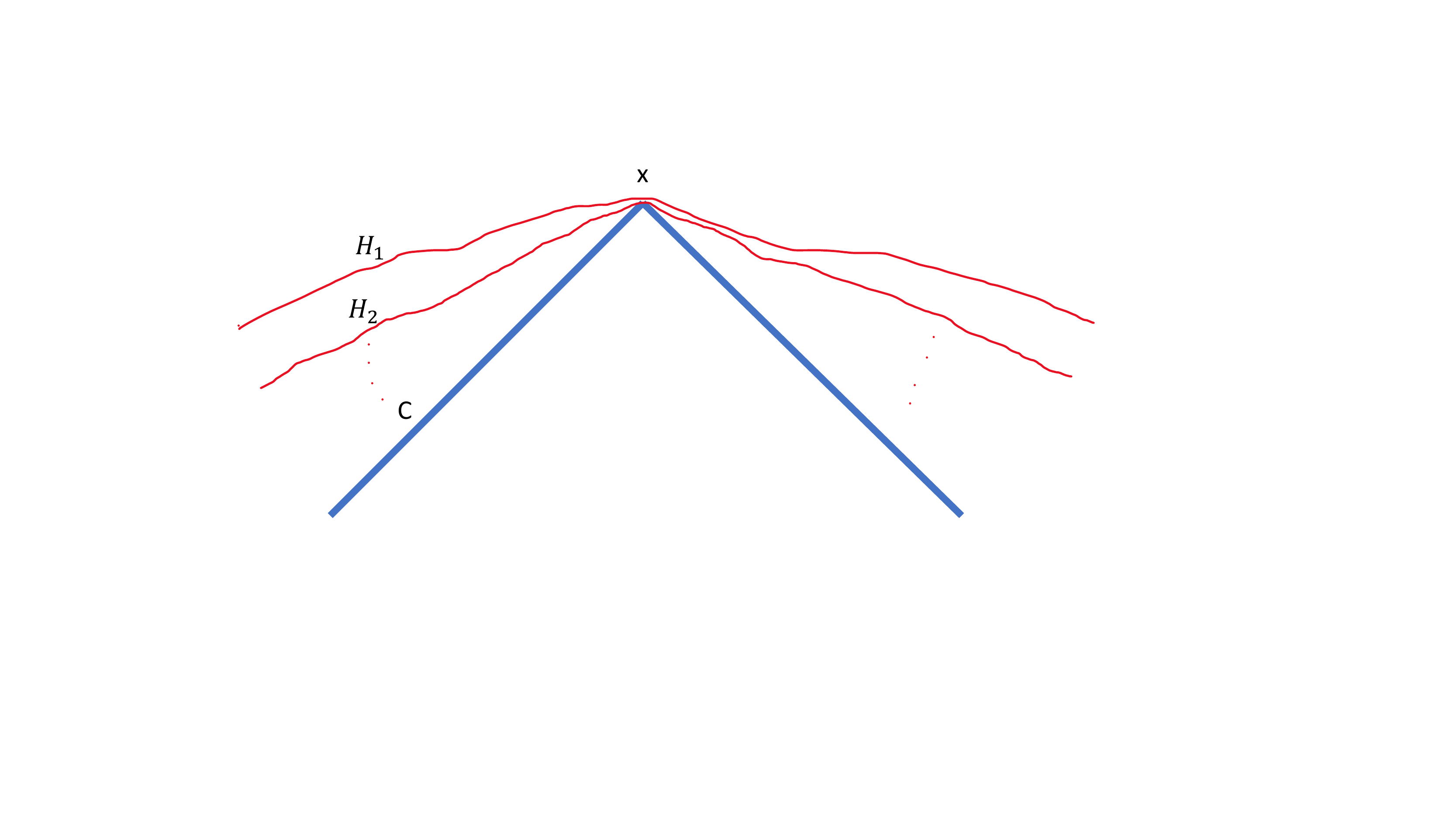}
\caption{Spacelike hypersurfaces tending to the past light cone}
\end{figure*}

Here $C$ is the surface of the past light cone of $x$. 
Take a family $\{ H_n : n = 1, 2, \ldots \}$ of spacelike 
hypersurfaces which go through $x$ and which
asymptotically tend to $C$.  
Write $H_n \setminus \{ x \}$ for the hypersurface $H_n$ 
with the point $x$ removed. 

Let $\ket { \psi^n }$ be the 
wave function of the full quantum system on $H_n$, and define
\begin{equation}\rho^n (x) = \Tr_{H_n \setminus \{ x \}} ( \ket{\psi^n } \bra{\psi^n } )
\, . 
\end{equation}
Then the local state at $x$ is
\begin{equation} 
\rho^{\rm loc} (x) = \lim_{ n \rightarrow \infty } \rho^n (x) \, . 
\end{equation}
Thus the local state is given by taking the joint wave
function of the complete system, defined by allowing for only
collapses in the past light cone of the 
particle, and then tracing out the rest of the system. 

To understand the essential point of this definition, consider a simplified 
example in which at time $t=0$ there are two entangled pointlike particles at
rest at spatial locations $L$ and $R$ in a state
\begin{equation}
a \ket{0}_L \ket{1}_R + b \ket{1}_L \ket{0}_R \, ,
\end{equation}
with zero Hamiltonian (in the absence of measurements).
Let the separation between $L$ and $R$ in their mutual rest frame be $d$.
At $t=0$, the local state at $L$ is then
\begin{equation}
| a |^2 \ket{0}_L \bra{0}_L + | b |^2 \ket{1}_L \bra{1}_L \, ;
\end{equation}
the local state at $R$ is
\begin{equation}
| b |^2 \ket{0}_R \bra{0}_R + | a |^2 \ket{1}_R \bra{1}_R \, .
\end{equation}
Here $\ket{0}, \ket{1}$ are orthonormal basis states of some 
physical qubit on each side. 
Now suppose that at $t=T$ a measurement is carried out in 
the  $\ket{0}_L, \ket{1}_L$ basis on the $L$ particle,
with outcome $0$.  At this point, the local state 
at $L$ becomes
\begin{equation}
 \ket{0}_L \bra{0}_L \, . 
\end{equation}
However, for $T \leq t < T+ \frac{d}{c}$, the local state
at $R$ remains 
\begin{equation}
| b |^2 \ket{0}_R \bra{0}_R + | a |^2 \ket{1}_R \bra{1}_R \, .
\end{equation}
This changes only when $t \geq  T+ \frac{d}{c}$, when a light signal 
describing the outcome of the $L$ measurement could have 
reached $R$: the local state then becomes
\begin{equation}
| a |^2 \ket{1}_R \bra{1}_R \, .
\end{equation}
Note that this definition holds whether or not any communication
actually occurs between observers at $L$ and $R$.
Indeed, we do not necessarily need to assume there {\it are} any
observers at $L$ and $R$, only that we are working with some
version of quantum theory in which collapse or measurement
events are defined and localised, so that it makes sense to
say that the $L$ particle is measured at $t=0$.   

In a more realistic model, the relevant systems are not pointlike
particles but have wave functions with extended support, as 
in Eqn. (\ref{ehepr}), or multi-particle wave functions, or 
field functionals.   
If classical gravitational fields interact with quantum
states in such a theory, it is natural to postulate that
they interact with $\rho^{\rm loc}(x)$, which also generally has extended
support at any given time.   Such interactions allow descriptions
of the quantum state of a localized system, in the sense that 
they could allow a complete classical readout of $\rho^{\rm loc}(x)$ 
for $x$ in a localized region of space-time, without causing 
any collapse.   A simulation argument shows that an effective readout device
of this type would not allow superluminal signalling \cite{kent2005nonlinearity}. 

\section{Conclusions}

No-go theorems in physics are often presented as insuperable
obstacles when they are actually only challenges to the imagination.   
This is the case with Eppley
and  Hannah's argument, which is based on a limited and
self-inconsistent model of the possible form of interactions 
between classical fields and quantum states.  

It was shown some time ago  \cite{kent2005nonlinearity}
that perfect classical state 
readout devices can be consistently added to relativistic quantum
theory without superluminal signalling, so long as the readout
devices produce a classical readout of the local state $\rho(x)$.
The implications of this result for non-standard theories 
and tests of gravity have been widely appreciated in the quantum
foundations community; for example, it plays a significant 
part in the Space QUEST mission proposal \cite{joshi2018space}.
However, they seem to have been less widely appreciated
in the quantum gravity community, and as far as I am aware the 
implications for the Eppley-Hannah argument have not
previously been discussed.   I hope this paper may help lay that argument to rest,
and lead to a broader understanding of the possibilities for classical-quantum
hybrid theories and a more balanced and nuanced view of the arguments for quantum
gravity.  

\vskip10pt
\begin{acknowledgments}
This work was partially 
supported by Perimeter Institute for Theoretical Physics. Research at Perimeter
Institute is supported by the Government of Canada through Industry
Canada and by the Province of Ontario through the Ministry of
Research and Innovation.
\end{acknowledgments}

\bibliographystyle{unsrtnat}
\bibliography{collapselocexpt}{}
\end{document}